Review

# Estimating Hubble Constant with Gravitational Observations: A Concise Review

Rosa Poggiani

Department of Physics, University of Pisa, 56127 Pisa, Italy; rosa.poggiani@unipi.it; Tel.: +39-050-2214432

**Abstract:** The Hubble constant is of paramount importance in astrophysics and cosmology. A large number of methods have been developed with different electromagnetic probes to estimate its value. The most recent results show a tension between values obtained from Cosmic Microwave Background observations and supernovae. The simultaneous detection of gravitational waves and electromagnetic radiation from GW170817 provided a direct estimation of the Hubble constant that did not depend on the astronomical distance ladder. This concise review will present the methods to estimate the Hubble constant with the gravitational observations of compact binary mergers, discussing both bright and dark sirens and reporting the state of the art of the results.

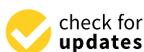

## 1. Introduction

The model of the Universe envisaged by Einstein was static and required the addition of a cosmological constant [1]. Friedmann found the solution for an expanding Universe [2], whose expansion rate was first estimated by Lemaitre [3]. The landmark observations and the estimated distances of 24 galaxies led Hubble to propose a relation between the recession speed $v$ and the distance $d$ [4], whose proportionality constant $H_0$ was later named the Hubble constant:

$$v = H_0 d \qquad (1)$$

The estimated value of the Hubble constant was approximately 500 km s$^{-1}$ Mpc$^{-1}$, compared with the currently accepted value of about 70 km s$^{-1}$ Mpc$^{-1}$.

The analysis by [5] of the Hubble data found that some stars in the sample were bright HII regions and the population of Cepheid stars was not homogeneous, estimating a value of the Hubble constant of 75 km s$^{-1}$ Mpc$^{-1}$.

Presently, there is a broad range of techniques to estimate the Hubble constant. Two reference values are based on electromagnetic observations but in different domains. The Planck Collaboration estimation, $H_0 = 67.4 \pm 0.5$ km s$^{-1}$ Mpc$^{-1}$, is based in the Cosmic Microwave Background (CMB) temperature and polarization anisotropies [6], while the Supernova H0 for the Equation of State (SH0ES) Collaboration value, $H_0 = 73.2 \pm 1.3$ km s$^{-1}$ Mpc$^{-1}$, is based in Cepheids and type Ia supernovae [7]. The inconsistence between the two estimates produces the Hubble tension. Due to the relevance of the Hubble constant in distance determinations and its relations with other cosmological parameters, it is of paramount importance to tackle the Hubble tension.

Extensive reviews discussing the methods and the observations to determine the Hubble constant have been presented by [8–11]. Figure 1 reports an overview of the estimates of the Hubble constant in literature secured using different methods, both in the electromagnetic domain and with gravitational waves, where the green and orange bands are the results by [6] and [7], respectively.

In addition to the CMB/Baryonic Acoustic Oscillations (BAO) estimates [6,12–15] and SN IA/Cepheids [7,16–19], a broad range of different methods are used: Tip of the Red Giant Branch (TRGB) with SN Ia [20–23]; Masers [24]; Tully–Fisher [25,26]; Surface Brightness Fluctuations (SBF) [27,28]; Miras [29]; SN II [30]; Time Delay Lensing (TDL) [31–33]; EBL gamma ray attenuation [34]; and GRB Fundamental Plane [35].

Generally, Hubble constant estimates in the electromagnetic domain rely on the observation of low-energy radiation (radio, optical, etc.). The EBL gamma ray attenuation and the GRB Fundamental Plane use high-energy probes as cosmological tools. In particular, the GRB Fundamental Plane uses GRBs as high redshift probes, with the three-dimensional Dainotti relation between the peak prompt luminosity, the rest frame time at the end of the X-ray plateau, and the corresponding X-ray luminosity [35–61].

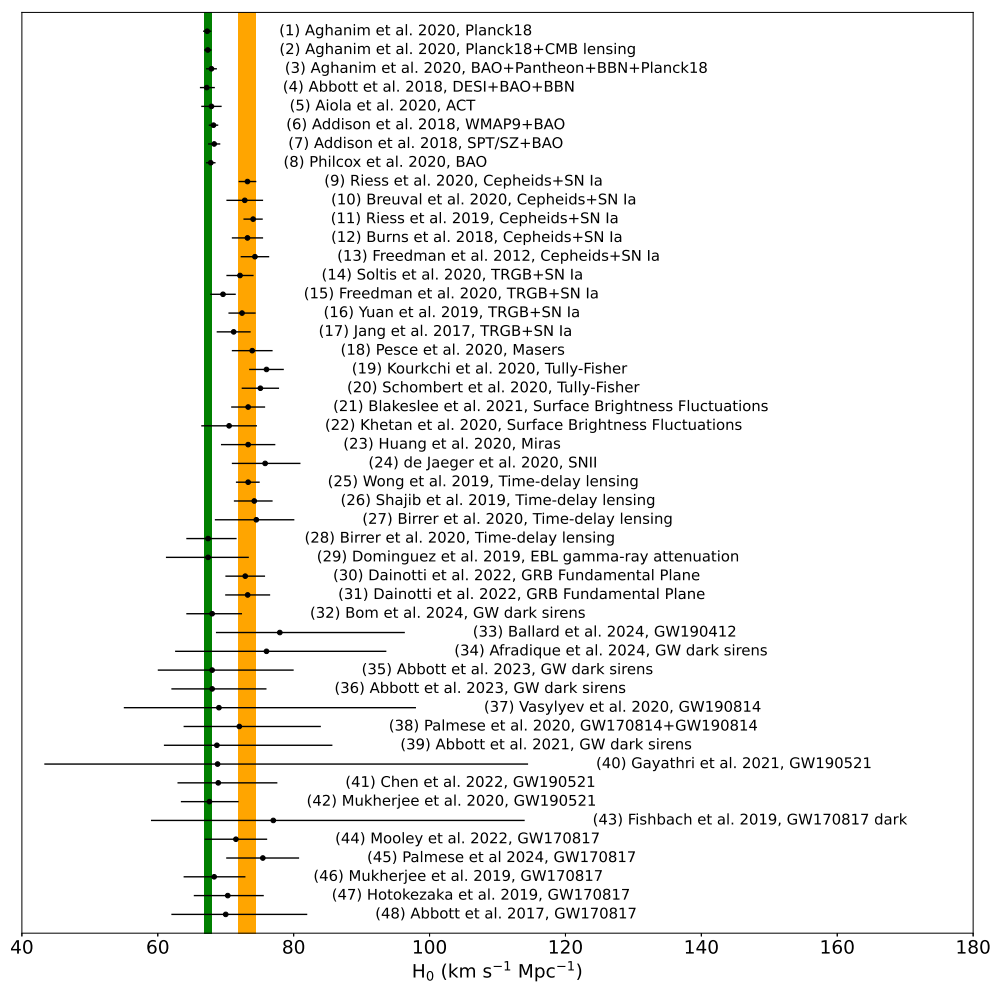

**Figure 1.** The value of Hubble measured through different methods in the literature. Planck CMB determinations (1), (2) (3) by [6]; Baryonic Acoustic Oscillations (BAO): (4) by [12]; CMB (5) by [13]; CMB + BAO: (6), (7) by [14]; (8) by [15]; Cepheids+SN IA (9) by [7]; (10) by [16]; (11) by [17]; (12) by [18]; (13) by [19]; Tip of the Red Giant Branch (TRGB)+SN Ia: (14) by [20]; (15) by [21]; (16) by [22]; (17) by [23]; Masers: (18) by [24]; Tully–Fisher: (19) by [25]; (20) by [26]; Surface Brightness Fluctuations (SBF): (21) by [27]; (22) by [28]; Miras: (23) by [29]; SN II: (24) by [30]; Time Delay Lensing (TDL): (25) by [31]; (26) by [32]; (27), (28) by [33]; EBL gamma ray attenuation: (29) by [34]; GRB Fundamental Plane: (30), (31) by [35]. The Hubble constant estimates with gravitational waves from (32) to (48) are discussed in detail in Sections 3, 4. The green and orange bands are the CMB results by [6] and the SN Ia/Cepheids results by [7], respectively.

With a few exceptions, such as masers, Hubble constant estimates rely on the astronomical distance ladder. Gravitational wave observations [62,63] can provide distance estimates completely independent from the electromagnetic observations, using standard sirens [64], the gravitational equivalent of the standard candles; see also [65].

The method relies on the relation between the strain amplitude h of the gravitational wave and the luminosity distance $D_L$:

$$h = \frac{2G}{c^4} \frac{1}{D_L} \ddot{I} \qquad (2)$$

where G is the gravitational constant and I is the mass quadrupole moment of the source. The strain h and the frequency evolution of the gravitational signal both depend on the component masses $m_1$, $m_2$ through the chirp mass $\mathcal{M}_c = (m_1 m_2)^{3/5}(m_1 + m_2)^{-1/5}$. The luminosity distance $D_L$ is estimated by measuring both the strain and the frequency evolution. Further details about distance measurements in gravitational astronomy can be found in [66].

Standard sirens can be bright, when an electromagnetic counterpart is observed and the host galaxy is identified, or dark. A first gravitational estimate of the Hubble constant [67] was performed using the GW170817 binary neutron star merger associated with the short Gamma-Ray Burst GRB 170817A in the host galaxy NGC 4993 [68]. Other Hubble constant estimates have been performed with dark sirens, relying on galaxy catalogs to provide a list of candidate host galaxies. The standard sirens will be discussed in Section 2; the Hubble constant determinations with bright sirens, in particular with GW170817/GRB 170817A, and with dark sirens will be discussed in Sections 3 and 4, respectively. Gravitational distance estimates have an impact not only in the context of General Relativity but also in the field of alternative theories of gravity [69–75]. The comparison of the luminosity distance for gravitational waves and for electromagnetic signals is also relevant for alternative models of gravity and the Universe [76,77].

## 2. Standard Sirens

The observation of gravitational waves has opened a new window to study the Universe, detecting mergers of binary black holes (BBH) [78], binary neutron stars (BNSs) [68], and neutron star black holes (NSBH) [79]. The most recent gravitational wave transient catalog, GWTC-3 [80], contains 90 compact binary mergers that have been used for population studies [81] and tests of General Relativity [82]. The gravitational observations alone allow a direct measure of the luminosity distance $D_L$ of the source, but not of the redshift, which must be separately estimated.

The method of bright sirens is based on the identification of the host galaxy of the merger, but the typical localization regions range from hundreds to thousands of deg$^2$, making the identification complicated unless an electromagnetic counterpart is detected. When the host galaxy is identified, its redshift can be determined from spectroscopic observations. An electromagnetic counterpart is expected for BNS and (at least some) NSBH mergers and has been observed in the GW 170817 BNS merger with the associated GRB 170817A [68], as discussed below; on the other hand, no counterpart is expected for BBH mergers, unless surrounding material is present in the merger region [83–89]. To date, GW170817 is the only merger with an unambiguous detection of an electromagnetic counterpart.

While gravitational wave observations alone do not provide information on the redshift z, they depend on the redshifted mass (1 + z)M, where M is the gravitational mass of the system. Therefore, the redshift can be inferred if an independent mass estimate is provided. The knowledge of either the mass distribution or the merger rate can lift the degeneracy between the mass and the redshift [90,91].

The method can be applied even when the electromagnetic counterpart is missing and has been used in the GWTC-3 estimation of the Hubble constant [92] described below.

When the electromagnetic counterpart of a merger is missing, the Hubble constant can be determined with dark sirens, combining the distance from gravitational wave observations with galaxy surveys to identify a list of candidate hosts within the localization volume, as first proposed by Schutz [64]. The method finds a general application to all types of mergers, including either black holes or neutron stars, BBHs, BNSs, and NSBHs. In addition to the large number of potential candidate galaxies within the large gravitational localization region, the method is limited by the completeness of the catalogs, which are flux-limited [93–95]. The dark siren method has been applied to the events detected in the O2 and O3 runs [92,96], as discussed below.

The compilation of estimates of the Hubble constant using bright and dark sirens, which will be discussed below, is reported in Table 1.

**Table 1.** Gravitational wave estimates of the Hubble constant.

| Method | $H_0$ (km s$^{-1}$ Mpc$^{-1}$) | Reference |
| --- | --- | --- |
| GW170817 bright siren | $70^{+12}_{-8}$ | [67] |
| GW170817 jet | $70.3^{+5.3}_{-5.0}$ | [97] |
| GW170817 jet | $68.3^{+4.6}_{-4.5}$ | [98] |
| GW170817 afterglow | $75.46^{+5.34}_{-5.39}$ | [99] |
| GW170817 jet | $71.5 \pm 4.6$ | [100] |
| GW170817 dark siren | $77^{+37}_{-18}$ | [101] |
| GW190521 | $67.6^{+4.3}_{-4.2}$ | [102] |
| GW190521 | $68.9^{+8.7}_{-6.0}$ | [103] |
| GW190521 | $68.8^{+45.7}_{-25.5}$ | [104] |
| O1, O2 dark sirens | $68.7^{+17.0}_{-7.8}$ | [96] |
| GW170814+GW190814 | $72.0^{+12}_{-8.2}$ | [105] |
| GW190814 | $69^{+29}_{-14}$ | [106] |
| O3 dark sirens | $68^{+8}_{-6}$ | [92] |
| O3 dark sirens, no catalogs | $68^{+12}_{-8}$ | [92] |
| Dark sirens | $76.00^{+17.64}_{-13.45}$ | [107] |
| GW190412 | $77.96^{+18.4}_{-9.4}$ | [108] |
| Dark sirens | $68.0^{+4.4}_{-3.8}$ | [109] |

## 3. Hubble Constant Measurements with Bright Sirens

*3.1. GW170817/GRB 170817A: Gravitational and Multi-Messenger Observations*

On 17 August 2017, the Advanced LIGO and Advanced Virgo interferometers detected the binary neutron star (BNS) merger GW170817 in the host galaxy NGC 4993 [68] (Figure 2). The luminosity distance of $40^{+8}_{-14}$ Mpc was consistent with the distance of NGC 4993 estimated with other methods [110–114]. The GBM instrument onboard Fermi and the SPI-ACS instrument onboard INTEGRAL detected the short Gamma-Ray Burst GRB 170817A after about 1.7 s at a position consistent with the merger [115,116]. The merger was detected with a combined signal-to-noise ratio of 32.4 and a false alarm rate smaller than one per $8.0 \times 10^4$ years in the LIGO Hanford and LIGO Livingston interferometers [68], and a signal-to-noise ratio of about 2 in the Virgo interferometer, indicating that the sky position of the merger was close to the region of null response [68]. The gravitational observations suggested that the range of the masses of the system components, 0.86 to 2.26 $M_\odot$, and the range of the total mass, 2.72 to 3.29 $M_\odot$, were consistent with the range of masses of neutron stars in binary systems [68].

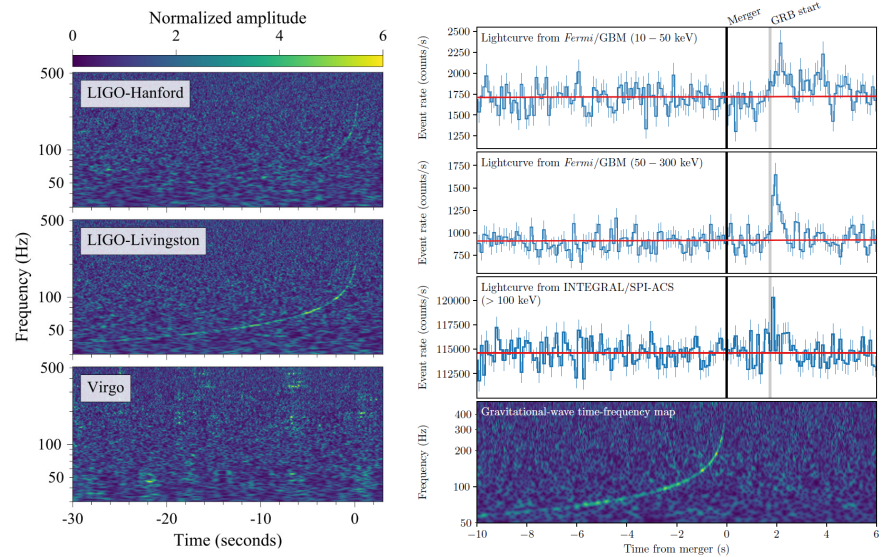

**Figure 2.** (**Left panel**): time−frequency maps of BNS merger GW170817 observed in the LIGO Hanford (**top**), LIGO Livingston (**center**), and Virgo (**bottom**) interferometers. (**Right panel**): signal of GRB 170817A, associated with GW170817, observed by the Fermi-GBM (10–50 keV, 50–300 keV) and INTEGRAL SPI-ACS instruments (the red line is the background estimate), and the time-frequency map of GW170817. Adapted from [68,117].

The 90% localization region was about 28 deg$^2$ [117], with a list of 54 galaxies obtained by cross-matching a galaxy catalog with the 90% localization volume of the merger; catalog incompleteness should be taken into account [118]. The optical counterpart of GW170817/GRB 170817A, AT2017gfo, was detected in the elliptical galaxy NGC 4993 10.87 h after the BNS merger by the One-Meter Two-Hemispheres collaboration using the Swope telescope [119,120] and quickly confirmed by DLT40 [121], VISTA [122], MASTER [123], DES [124], and LCO [125]. GW170817 was localized at a projected distance of about 2 kpc from the galaxy center [126], an offset consistent with the offsets of short GRBs (sGRBs) in other galaxies [127,128]. The images secured four months before the merger did not show any candidates [121]. The detection of associated electromagnetic radiation and the mass range of both primary and secondary show that at least one of the merging objects was a neutron star. The extensive follow-up multi-messenger observations of GW170817/GRB 170817A around the epoch of the merger and in the first days after have been summarized by [129]. No X-ray or radio counterpart was detected during the first week after the merger [116,130–134]. The X-ray and radio afterglows emerged 9 days [135] and 16 days [132] after the merger, respectively. Additional details about GW170817 and its afterglow can be found in a companion paper in this Special Issue.

*3.2. GW170817/GRB 170817A: Hubble Constant Measurement with the Electromagnetic Counterpart*

NGC 4993 shows a bulge, concentric shells, and dust lanes, and contains an old stellar population with a low Star Formation Rate [136]. Assuming that the optical counterpart defines the effective sky position of the merger, a re-analysis of gravitational data alone leads to a distance of $43.8^{+2.9}_{-6.9}$ Mpc [67]. In addition to calibration uncertainties and noise in the interferometers, the uncertainty on the distance is produced by the degeneracy between the luminosity distance $D_L$ and the orbital inclination angle $i$ [137]. The distance and the orbital inclination contribute as $(1 + \cos^2 i)/D_L$ and $\cos i/D_L$ for the two gravitational wave polarizations. A similar signal amplitude can be produced either by a close binary with high inclination or by a distant face off/on merger. Degeneracy occurs when only a single polarization amplitude is measured [137]. The two LIGO interferometers are almost aligned and sensitive to a single polarizations; thus, an additional detector such as Virgo is required to measure both polarizations. The uncertainty has been later removed with the detection of the kilonova [122,138] and the presence of a collimated jet [139], as will be discussed in detail in Section 3.2.

The detection of the merger and its associated GRB has triggered new measurements of the distance to NGC 4993 using the Fundamental Plane (FP), Surface Brightness Fluctuations (SBFs), and Globular Cluster Luminosity Function (GCLF) [111–114], as reported in Table 2. Previous measurements were based on the Tully–Fisher method, with distances in the range 31.0 to 41.1 Mpc [110,140–142]. The gravitational wave distance estimate is consistent with the distances to NGC 4993 measured with the other methods.

**Table 2.** Distance to NGC 4993 using different methods.

| Method | Distance (Mpc) | Reference |
| --- | --- | --- |
| Fundamental Plane | $41.0 \pm 3.1$ | [112] |
| Fundamental Plane | $37.7 \pm 8.7$ | [113] |
| Surface Brightness Fluctuations | $40.7 \pm 1.4 \pm 1.9$ | [111] |
| Globular Cluster Luminosity Function | $41.65 \pm 3.00$ | [114] |

The determination of $H_0$ [67] requires the estimation of the background Hubble flow velocity at the galaxy position. NGC 4993 belongs to a group of galaxies whose center of mass recession velocity relative to the frame of the Cosmic Microwave Background (CMB) is $3327 \pm 72$ km s$^{-1}$ [143,144]. The group velocity must be corrected by the peculiar velocity of 310 km s$^{-1}$ due to the bulk motion toward the Great Attractor [145,146]. A conservative estimate of 150 km s$^{-1}$ has been assumed as the uncertainty on the peculiar velocity at the location of NGC 4993 [146]. The estimated Hubble velocity is $3017 \pm 166$ km s$^{-1}$ [67]. Since the gravitational measurement of the distance is correlated with the inclination of the orbital plane of the binary and on the properties of its components, the uncertainty in these variables is managed by marginalizing over the posterior distribution on system parameters [67], considering a fixed position in the sky at the location of the optical counterpart. A Bayesian analysis to infer a posterior distribution on $H_0$ and system inclination, with marginalization over uncertainties in the peculiar and recessional velocity, produced the marginal posterior for $H_0$ reported in Figure 3.

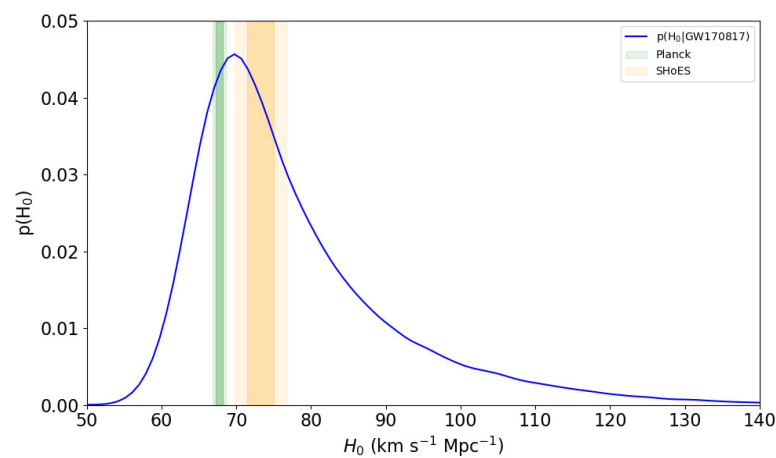

**Figure 3.** Hubble constant posterior from GW170817 [67]. The shaded regions are the 1$\sigma$ and 2$\sigma$ contours of the Planck CMB [147] and (SH0ES) supernova [148] estimates. Data credits: https://dcc.ligo.org/LIGO-P1700296, accessed on 22 May 2025.

The inferred value of $H_0$ was $70^{+12}_{-8}$ km s$^{-1}$ Mpc$^{-1}$ [67], consistent with other estimates: the Planck CMB measurements ($67.74 \pm 0.46$ km s$^{-1}$ Mpc$^{-1}$) [147]; the type Ia supernova measurements from SH0ES (73.24 ± 1.74 km s$^{-1}$ Mpc$^{-1}$) [148]; the Cepheid measurements from the Hubble Space Telescope Key Project [110]; the $H_0$ Lenses in COSMOGRAIL's Wellspring (H0LiCOW) strong lensing measurements [149]; the SDSS Baryon Acoustic Oscillations measurements [150]; and the SPT high angular multipole CMB measurements from SPT32 [151]. The gravitational estimation of $H_0$ was broadly consistent with both the Planck and SH0ES results but did not resolve the Hubble tension.

The Very Long Baseline Interferometry (VLBI) follow-up of the electromagnetic afterglow of GW170817 provided a tool to resolve the degeneracy of distance and inclination, revealing the presence of a collimated jet [139]. The detection of the jet triggered new estimations of the Hubble constant based on the modeling of the jet afterglow light curve and on the centroid motion. These methods have systematic uncertainties arising from modeling uncertainties in the jet morphology and from the choice of parameter priors. The estimation of the Hubble constant by [97], $H_0 = 70.3^{+5.3}_{-5.0}$ km s$^{-1}$ Mpc$^{-1}$, used the superluminal motion of the jet and also provided a measurement of the orbital inclination under the assumption of jet emission orthogonal to the orbital plane of the merger. The combination of the VLBI observations with a new estimation of the peculiar velocity produced a revised value of $H_0 = 68.3^{+4.6}_{-4.5}$ km s$^{-1}$ Mpc$^{-1}$ [98]. The later observations of the GW170817 afterglow have been used for additional Hubble constant estimates. The optical, radio, and X-ray observation of the jet afterglow up to 3.5 years after the merger has produced a measure of the viewing angle of the binary, $30.4^{+2.9}_{-1.7}$ deg, breaking the distance-angle degeneracy, and an estimation of the Hubble constant as $75.46^{+5.34}_{-5.39}$ km s$^{-1}$ Mpc$^{-1}$ [99]. A more recent determination of the Hubble constant [100], 71.5 ± 4.6 km s$^{-1}$ Mpc$^{-1}$, relies on the optical superluminal motion measurements at different epochs.

The GW170817 event has also been analyzed as a dark siren without its electromagnetic counterpart using the Galaxy List for the Advanced Detector Era (GLADE catalog) [101], with an estimated $H_0 = 77^{+37}_{-18}$ km s$^{-1}$ Mpc$^{-1}$.

The number of bright sirens is expected to increase in the future, allowing us to infer the redshift by observing the electromagnetic counterpart [65,152–154]. The initial predicted estimate of the Hubble constant at a level of about two per cent within five years suggested in 2018 by [153] has, however, already been falsified by observations. The number of required mergers can decrease if electromagnetic observations provide constraints on the system inclination as in the observations described above.

*3.3. Bright Sirens Other than GW170817?*

Among the 90 mergers detected in the first three observing runs [80], there is a second binary neutron star merger, GW190425 [155], with component masses of $2.1^{+0.5}_{-0.4}$ M$_\odot$ and $1.3^{+0.3}_{-0.2}$ M$_\odot$ [156]. The large luminosity distance, $0.15^{+0.08}_{-0.06}$ Gpc, and the extended localization region, 8700 deg$^2$, of GW190425 are less favorable than those of GW170817. Despite an extensive follow-up campaign [157–159], there is no accepted counterpart. However, the Anti-Coincidence Shield (ACS) of the SPI gamma-ray spectrometer of INTEGRAL detected GRB 190425 [160], but the observation was not confirmed by Fermi-GBM [161]. The CHIME observatory detected the Fast Radio Burst FRB 20190425A about 2.5 h after the merger [162] at a position consistent with UGC 10667 [163], but the possible association with the merger was ruled out by the lack of a kilonova counterpart [164]. No counterpart was detected for the two neutron-star-black-hole mergers, GW200105 and GW200115 [79].

A special case of the possible bright siren is GW190521, a BBH merger with a total mass of about 150 M$_\odot$ [165,166], that has been associated with the transient ZTF19abanrhr, a flare of the active galactic nucleus J124942.3+344929 [167] produced by the interaction of the remnant black hole with the accretion disk [88]. There is no consensus about the association of the merger with the flare [168,169]. GW190521 has been used alone or by adding the GW170817 posterior to estimate the Hubble constant: $67.6^{+4.3}_{-4.2}$ km s$^{-1}$ Mpc$^{-1}$ [102]; $68.9^{+8.7}_{-6.0}$ km s$^{-1}$ Mpc$^{-1}$ [103]. A value of Hubble constant of $68.8^{+45.7}_{-25.5}$ km s$^{-1}$ Mpc$^{-1}$ has been derived by [104] in the assumption of an eccentric merger.

## 4. Hubble Constant Measurements with Dark Sirens

A first estimation of the Hubble constant with dark sirens used BBH mergers detected during the O1 and O2 runs [96] combined with the GLADE catalog of galaxies observed in the B band [170]. The combined estimate using a set of six BBH mergers without a counterpart and the GW170817 estimate produced $H_0 = 68.7^{+17.0}_{-7.8}$ km s$^{-1}$ Mpc$^{-1}$ [96]. Several recent estimates of the Hubble constant have used the mergers of the O3 run. The estimation based on the GWTC-3 catalog used 47 compact binary coalescences detected during the O1, O2, and O3 runs and the updated GLADE+ catalog, which is complete up to a luminosity distance of $47^{+4}_{-2}$ Mpc [171]. The majority of mergers routinely detected are at large distances, where the GLADE+ catalog is incomplete. While the dark sirens alone lead to $H_0 = 67^{+13}_{-12}$ km s$^{-1}$ Mpc$^{-1}$, the inclusion of GW170817 produces an improved value, $H_0 = 68^{+8}_{-6}$ km s$^{-1}$ Mpc$^{-1}$, also in comparison to the original GW170817 estimation [92].

Additional estimates of the Hubble constant have been performed with one or a selection of dark sirens and catalogs with completeness extending to larger distances than GLADE+. The dark sirens GW170814 and GW190814 have been used together with the Dark Energy Survey (DES) catalog, producing $H_0 = 75^{+40}_{-32}$ km s$^{-1}$ Mpc$^{-1}$ [172] and $78^{+57}_{-13}$ km s$^{-1}$ Mpc$^{-1}$ [105], while combining GW170814, GW190814, and GW170817 leads to $H_0 = 72.0^{+12}_{-8.2}$ km s$^{-1}$ Mpc$^{-1}$ [105]. GW190814 and the GLADE catalog have been used by [106] to derive $H_0 = 69^{+29}_{-14}$ km s$^{-1}$ Mpc$^{-1}$ in combination with the GW170817 posterior. An estimation has used the merger GW190412 [173], which is lacking an electromagnetic counterpart but is localized within about 12 deg$^2$ as a dark siren [174]. The estimated Hubble constant is $85.4^{+29.1}_{-33.9}$ km s$^{-1}$ Mpc$^{-1}$, which becomes $77.96^{+18.4}_{-9.4}$ km s$^{-1}$ Mpc$^{-1}$ when the GW170817 posterior is included [108]. Using GW190924_021846 and GW200202154313 as standard sirens and the DECam Local Volume Exploration (DELVE) catalog together with the bright siren GW170817 produces $H_0 = 68.84^{+15.51}_{-7.74}$ km s$^{-1}$ Mpc$^{-1}$, while using 10 well-localized (within 12 to ∼400 deg$^2$) O3 dark sirens leads to $H_0 = 76.00^{+17.64}_{-13.45}$ km s$^{-1}$ Mpc$^{-1}$ [107]. The mergers detected during the O4 run have triggered a new determination of Hubble constant, combining 5 new dark sirens with 10 dark sirens from the previous runs [109], with $H_0 = 70.4^{+13.6}_{-11.7}$ km s$^{-1}$ Mpc$^{-1}$ and $68.0^{+4.4}_{-3.8}$ km s$^{-1}$ Mpc$^{-1}$ when combined with the GW170817 posterior.

All estimates based on dark sirens are broadly consistent and also consistent with the GW170817 estimate. The results suggest that well-localized dark sirens and complete galaxy catalogs can approach the precision of the bright sirens.

When using dark sirens, the modeling of the underlying mass distribution has a strong impact on the Hubble constant, and in principle, it is possible to obtain the BBH mass distribution together with the Hubble constant [175,176]. The Hubble estimation using the analysis of the first two runs used a simple power law model [96]. The Hubble constant determination based on the events in the first three observing runs investigated different distributions, in particular, a power law with one Gaussian over-density at about 35 M$_\odot$ [92]. An analysis not relying on galaxy catalogs provided a joint constraint on the properties of the population of BBHs and the parameters of the cosmological model, with an estimated $H_0 = 68^{+12}_{-8}$ km s$^{-1}$ Mpc$^{-1}$ [92].

Additional methods have been suggesting to estimate the redshift of dark sirens: redshift distribution of the gravitational sources [177,178]; cross-correlation techniques for deriving the clustering redshift of gravitational mergers [179,180]; Pair-Instability Supernovae mass scale of black holes [91,175]; and tidal distortion of neutron stars [181,182].

## 5. Conclusions

Gravitational observations of mergers provide a tool to estimate the Hubble constant. The paper has summarized the estimates based on both bright sirens and dark sirens. The results from the bright siren GW170817 and the associated electromagnetic counterpart have provided the first estimation of the Hubble constant independent of the astronomical distance ladder. Other estimates use dark sirens, possibly well localized, and catalogs, whose completeness is a key factor. The number of detected mergers is steadily increasing, with the perspective of detecting a BNS merger as GW170817 in the near future and to estimate the redshift of dark sirens.

**Funding:** This research received no external funding.

**Acknowledgments:** The author is grateful to the referees for their useful comments that improved the paper.

**Conflicts of Interest:**